\documentclass[aps,twocolumn]{revtex4-1}
\usepackage{graphicx}
\usepackage{color}
\usepackage{ amssymb }
\usepackage{ amsmath }
\newcommand{\ket}[1]{\left| #1 \right\rangle}

\newcommand{\Tr}{\mathrm{Tr}}

\newcommand{\Prob}{\mathrm{P}}
\newcommand{\B}{{\cal B}}
\newcommand{\Bq}{{\cal B}_{\cal Q}}
\newcommand{\Bqo}{{\cal B}_{{\cal Q}1}}
\newcommand{\Bqz}{{\cal B}_{{\cal Q}0}}

\newcommand{\pool}{\mathrm{pool}}

\newcommand{\eea}{\end{eqnarray}}
\newcommand{\bea}{\begin{eqnarray}}
\newcommand{\ee}{\end{equation}}
\newcommand{\be}{\begin{equation}}
\newcommand{\ese}{\end{subequations}}
\newcommand{\bse}{\begin{subequations}}

\newcommand{\varT}{{\cal T}}

\newcommand{\unique}{{\rm unq}\,}

\newcommand{\shared}{{\rm red}\,}
\newcommand{\synergy}{{\rm syn}\,}
\newcommand{\syn}[2]{\synergy #1 \& #2}
\newcommand{\unq}[2]{\unique #1 \backslash #2}
\newcommand{\red}[2]{\shared #1 \& #2}
\newcommand{\setn}[1]{\{#1\}}
\newcommand{\bhat}{Bhattacharyya\,}

\begin{document}
\title{Quantum Partial Information Decomposition}
\author{S.J. van Enk}
\affiliation{Department of Physics and
	Oregon Center for Optical, Molecular \& Quantum Sciences\\
	University of Oregon, Eugene, OR 97403}
\begin{abstract}
The Partial Information Decomposition (PID)
takes one step beyond Shannon's theory in decomposing the information two variables $A,B$ possess about a third variable $\varT$ into distinct parts: 
unique, shared (or redundant) and synergistic information.
Here we show how these concepts can be defined in a quantum setting. We apply a quantum PID to scrambling in quantum many-body systems, for which a quantum-theoretic description has been proven productive.
Unique information in particular provides a finer description of scrambling than does the so-called tri-information.
\end{abstract}
\maketitle
\section{Introduction}
The  Partial Information Decomposition (PID) is a method 
for disentangling the relations between multiple random variables as encoded in their joint probabity distribution.
The method was conceived of in Ref.~\cite{williams2010} and in the simplest nontrivial case of three variables \footnote{For two variables Shannon's mutual information \cite{shannon1948,cover1999} suffices.} defines
synergistic information, unique information, and redundant information.
One important motivating example for desiring to go beyond Shannon's information theory is given in Ref.~\cite{james2017}.
There two different probability distributions $P_1$ and $P_2$  over three variables with distinct underlying mechanisms are presented, which cannot be distinguished by any of the standard quantities defined within Shannon's theory.
Specifically, mutual information between any combination of the three variables  cannot distinguish $P_1$ from $P_2$. Hence, linear combinations of mutual information quantities, such as co-information  \footnote{Co-information is defined as $I(\varT;A,B)-I(\varT;A)-I(\varT;B)$. It is also known as interaction information \cite{mcgill1954}, and has also been denoted as synergy \cite{brenner2000,schneidman2003,latham2005}. In terms of the PID, co-information equals the difference between synergistic and redundant information \cite{williams2010}.}  (whose negative is known in the quantum context as tri-information) can also not distinguish $P_1$ and $P_2$.
The PID, however, does distinguish these: all newly introduced quantities---synergistic, unique, and shared information---differ for $P_1$ and $P_2$. (See Section \ref{three} for the details.)

Using its ability to make distinctions between different probabilistic mechanisms, the PID  has been applied to descriptions and understanding of complex networks \cite{battiston2021,rosas2019}, and neural networks in particular \cite{tax2017,wibral2017a,wibral2017b,newman2022}.  For an overview of its uses, see \cite{lizier2018}. Several different proposals for a PID exist, based on different definitions of either unique information \cite{bertschinger2014,james2018b} or redundant information \cite{williams2010,harder2013,griffith2014,ince2017,finn2018,kolchinsky2022} or synergistic information \cite{rosas2020,vanenk2023}.
The idea that a finer distinction between different types of information  is useful had been around   in neuroscience \cite{brenner2000,schneidman2003,latham2005} before the seminal Ref.~\cite{williams2010}.

Since all of Shannon's classical concepts have been generalized to quantum settings, and since such generalizations have proven very fruitful  \cite{nielsen2010,hayashi2016,wilde2013,watrous2018},
there ought to be a quantum version of the PID, a QPID, as well. We define our version in Section III.
As we will show in Section IV, the classical motivating example distributions $P_1$ and $P_2$ can be quantized, such that standard quantum mutual information quantities (including tri-information) do not distinguish the two corresponding pure states $\ket{\Psi_1}$ and $\ket{\Psi_2}$. 
Just like the classical PID, the QPID is not unique, but the version proposed here does clearly distinguish these two pure states.  Moreover, when applying the PID  concepts to the issue of quantum scrambling \cite{hayden2007} in quantum many-body systems, the particular choice we make here gives sensible numerical results, especially for the unique information.
There is a good reason for expecting unique information to play a role in a quantum context. Whereas classical correlations can be shared unrestrictedly, the no-cloning theorem (or the monogamy of entanglement, see e.g., Ref.~\cite{dhar2017}) prohibits two systems to possess the same maximal quantum entanglement with a third system, thus typically forcing both systems to possess some unique quantum information about the third system.  (See Section IV.B for details.)

We start by summarizing some relevant aspects of the classical PID.
\section{PID for three variables}\label{three}
Consider two classical variables $A$ and $B$ correlated with a third classical variable $\varT$ that is the target of our inquiry.
We assume the joint probability distribution $\Prob(\varT,A,B)$ exists.
We may then define the Shannon entropy  \cite{shannon1948,cover1999}
\be
H(X)=-\sum_{\varT,A,B} \Prob(\varT,A,B) \log_2(\Prob(X)),
\ee
in which the symbol $X$ may stand for any subset of the variables $A,B,\varT$ as well as for any conditional
variable, such as $\varT|A$.  We use here logarithms base 2, so information and entropy will be given in units of bits.

How much information about $\varT$ 
we can obtain from variable $A$ alone is given by the mutual information \cite{shannon1948,cover1999}
\bse\label{MI}
\bea
I(\varT;A)&=&H(\varT)-H(\varT|A)\label{defasym}\\
&=&H(\varT)+H(A)-H(\varT,A)\label{defsym}.
\eea
\ese
That is, according to the first line, information equals the decrease in  entropy when going from an initial distribution $\Prob(\varT)$ to a final  probability distribution $\Prob(\varT|A)$. The second line shows the mutual information thus defined is actually symmetric between the two variables $A$ and $\varT$.

%\subsection{Motivating example}
An important motivation for extending Shannon's theory and going beyond mutual information, is the following example, taken from Ref.~\cite{james2017}. 
In Tables I and II we display two  different joint probability distributions for our three variables $\varT,A,B$, each taking on four possible values. 
These two distributions cannot be distinguished by any measures constructed from just the Shannon entropies of the above type.
\begin{table}[h]\label{C1}
	\begin{center}
		\begin{tabular}{ |c|c| } 
			\hline
			$\varT,A,B$ &		Probability  \\ 
			\hline
			$0,0,0$&		1/8  \\ 
			$0,2,2$&		1/8 \\ 
			$2,0,2$&		1/8  \\ 
			$2,2,0$&		1/8 \\ 
			$1,1,1$&		1/8  \\ 
			
			$1,3,3$& 1/8\\
			
			$3,1,3$&		1/8  \\ 
			$3,3,1$& 1/8\\
			\hline
		\end{tabular}

	\end{center}
	\caption{Probability distribution taken from Ref.~\cite{james2017} (the ``triadic'' case). 
		Each of the three variables can take on 4 values, 0,1,2,3.
		The information obtainable from $\Prob(\varT|A)$ is the same as that obtainable from $\Prob(\varT|B)$.  Namely, 
		when $A$ is even (odd), we can conclude $\varT$ is even (odd), but nothing more, and the same conclusion about $\varT$ follows from  $B$.
	}
\end{table}

\begin{table}[h]\label{C2}
	\begin{center}
		\begin{tabular}{ |c|c| } 
			\hline
			$\varT,A,B$ &		Probability  \\ 
			\hline
			$0,0,0$&		1/8  \\ 
			$0,2,1$&		1/8 \\ 
			$1,0,2$&		1/8  \\ 
			$1,2,3$&		1/8 \\ 
			$2,1,0$&		1/8  \\ 
			
			$2,3,1$& 1/8\\
			
			$3,1,2$&		1/8  \\ 
			$3,3,3$& 1/8\\
			\hline
		\end{tabular}
	\end{center}
	\caption{Probability distribution taken from Ref.~\cite{james2017} (the ``dyadic'' case), to be contrasted with the distribution from Table I. 
		Each of the three variables can take on 4 values, 0,1,2,3.
		The information obtainable from $\Prob(\varT|A)$ differs from that obtainable from $\Prob(\varT|B)$.  For example, from $A=0$ we can conclude that either $\varT=0$ or 1. From $B=0$ we can conclude $\varT=0$ or 2. There is a crucial bit of information we get uniquely from $A$ and another unique bit we get from $B$. Together, these fix the value of $\varT$ (in this example, we conclude $\varT=0$). 
	}
\end{table}

For example, each has a joint entropy $H(\varT,A,B)=3$,
and for each we have $I(\varT;A)=I(\varT;B)=1$. Using more general entropy functions, such as the  Renyi entropy $H_{\alpha}$ for $\alpha \geq 0$ \cite{renyi1961},  does not help either in distinguishing the two \cite{james2017}.
As the Table captions explain, there is a difference, though, in what sort of  information the individual variables $A$ and $B$ carry about $\varT$. Even though the amounts are the same (1 bit), in the first case the variables carry the same bit of information, in the second they carry different bits.
In other words, in the latter case, each variable carries some information that is unique w.r.t. the other variable.

We thus wish to quantify how much information about $\varT$ from $B$ is unique (relative to $A$, that is), and how much information is unique to $A$ (relative to $B$). 
In the context of the PID we frame this problem as follows. We try to write the three known standard mutual information quantities $I(\varT;A)$, $I(\varT;B)$ and $I(\varT;A,B)$ that concern information about $\varT$ as linear combinations of four new quantities (only one of which is, therefore, linearly independent of the 3 standard quantities). Two of these new quantities are meant to quantify the two types of unique information, the other two then give redundant and synergistic information, like so:
\bea \label{PID}
I(\varT;A)&=&I_{\unq{A}{B}} + I_{\red{A}{B} },\nonumber\\
I(\varT;B)&=&I_{\unq{B}{A}} + I_{\red{A}{B} },\nonumber\\
I(\varT;A,B)&=& I_{\unq{A}{B}} +I_{\unq{B}{A}} +  I_{\red{A}{B}} + I_{\syn{A}{B}}.\nonumber\\
\eea
Since we always consider information about the variable $\varT$ we suppress reference to it in the quantities appearing on the right-hand side.

Since we have introduced four quantities, but have only three equations for them, one more constraint is needed to define the PID.  Disagreement has arisen in the literature over what ought to be the fourth equation. Our view is that there is indeed a freedom of choice, and that distinct choices make sense in different contexts. Here we will make a choice that can  be straightforwardly generalized to a quantum setting (but there may well be other such choices \footnote{For example, Ref.~\cite{james2018b} defines a PID in terms of secret key rates. One expects ideas from quantum secret sharing \cite{hillery1999} to lead to a QPID, too.}). 

Here is one way to define unique information \cite{vanenk2023}.
We define for each pair of possible  values $a,b$ for the variables $A,B$ 
\be\label{ZZ}
Z_{ab}=\sum_t\Prob(t|a)^{1/2}\Prob(t|b)^{1/2},
\ee
where the sum is over the values $t$ the variable $\varT$ can take.
This is an overlap between two distribution functions, the \bhat measure \cite{bhat1946}.
It lies between 0 (for orthogonal distributions, which have no common support; this extreme case cannot occur in our setting) and 1 (for identical distributions).
The idea is that unique information only exists when
the conditional distributions $\Prob(t|a)$ and $\Prob(t|b)$ are different for at least some possible values of $a,b$.
We define the nonnegative quantity $\B_1$ by
\be\label{BB}
\B_1=-\sum_{a}\sum_{b} \Prob(a,b)\log_2(Z_{ab}),
\ee
where 
\be
\Prob(a,b)=\sum_t \Prob(t,a,b)
\ee
is the joint distribution for $a,b$.
An operational meaning for $\B_1$ is given in  Ref.~\cite{vanenk2023} in terms of pooling probability distributions \cite{lind1988,pool2000,carvalho2023,neyman2022}: given the two distributions $\Prob(t|a)$ and $\Prob(t|b)$ for fixed $a,b$, we can generate a single distribution over $t$ by choosing 
\be
P_\pool(t)=\Prob(t|a)^{1/2} \Prob(t|b)^{1/2}/Z_{ab},
\ee
where $Z_{ab}$ appears as a normalization factor. (This method of pooling is called ``logarithmic.'')
For this pooled distribution the average uncertainty about $\varT$ is reduced  relative to the average entropy $\tfrac12 [H(\varT|A)+H(\varT|B)]$ by an amount $\B_1$.  $\B_1$ can thus be considered a ``bonus'' amount of information.

An alternative way to define a single distribution with less uncertainty than this average amount is the trivial way, of choosing whichever of $\Prob(t|a)$ and $\Prob(t|b)$ has the lowest entropy on average (averaged over all values of $a,b$). For that trivial choice the reduction of uncertainty (an alternative  ``bonus'') would equal
\be
\B_0=\frac12 \left| H(\varT|A)-H(\varT|B)\right|.
\ee
We then define unique information by using the larger of the bonuses $\B_1$ and $\B_0$, 
\be\label{Bmax}
\B:=\max(\B_0,\B_1),
\ee
as 
\be
I_{\unq{A}{B}}+I_{\unq{B}{A}}=
2\B.
\ee
The individual unique information quantities then follow from (\ref{PID}),
\bea\label{unq}
I_{\unq{A}{B}}&=& \B +\frac12\left( H(\varT|B)-H(\varT|A) \right),\nonumber\\
I_{\unq{B}{A}}&=& \B +\frac12\left( H(\varT|A)-H(\varT|B)\right),
\eea
which are both nonnegative.
We note that with these definitions we can indeed clearly distinguish the two probability distributions $P_1$ and $P_2$ from Tables I and II: we find $\B=0$ for $P_1$ (and hence zero unique information) but $\B=\B_1=1$ (and hence 1 bit of unique information for both $A$ and $B$) for $P_2$, exactly agreeing with the intuition given in the Table captions.

We may view $\B$ as quantifying the asymmetry between $A$ and $B$ w.r.t.~their correlations with $\varT$. 
We may also view $\B$ as the fourth independent quantity, besides three mutual information measures, that characterizes how two variables may contain information about a third variable.

To conclude this Section on the classical PID, let us note that not all quantities named ``information'' in the decomposition (\ref{PID}) are differences between two entropies. In particular, for the decomposition based on (\ref{unq}), and similarly, for the decomposition proposed in \cite{bertschinger2014} the ``redundant information'' actually is a difference between two information quantities. As such, it has been denoted by $\Delta I_{\red{A}{B}}$ in early references \cite{latham2005} and \cite{schneidman2003}, as well as in \cite{vanenk2023}. Being a difference between two (non-negative) information quantities, $\Delta I_{\red{A}{B}}$ may take on negative values.  It follows directly from (\ref{PID}) that unique information is larger than the mutual information in such a case. The same will then be true of its quantum generalization.
\section{From classical to quantum information}
When trying to define quantum information quantities by replacing classical variables  by quantum systems, replacing probability distributions by density operators $\rho$, and replacing
the Shannon entropy function by the Von Neumann entropy
$S(\rho)=-\Tr \rho \log_2(\rho)$,
one runs into two well-known sorts of issues.
First, different density operators do not commute in general.  Second, unlike in the classical case, it matters whether we perform measurements
or not, and, in addition, what measurements we perform.  Both issues are illustrated below.

Quantum information theory \cite{nielsen2010,hayashi2016,wilde2013,watrous2018} has taught us how to circumvent such difficulties and define meaningful quantities that generalize classical quantities. Often, multiple generalizations exist for a given classical quantity (with different interpretations), for the reasons alluded to above. We will see that here, too, multiple possibilities exist to generalize the PID to a quantum version. For various reasons mentioned below, we do focus here on one particular generalization.

\subsection{Mutual information}
Given the two equivalent ways (\ref{defasym}) and (\ref{defsym}) in which we can define classical mutual information $I(\varT;A)$, there are two ways to generalize mutual information to a quantum setting. As is well-known, these two generalizations are no longer equivalent.

The first generalization is explicitly not symmetric between the two variables: we assume we perform a measurement on system $A$ and produce a density operator for system $\varT$ that depends on the measurement outcome.
We can always describe the measurement on $A$ by a set of POVM elements $\setn{\Pi_n}$ labeled by the outcome $n$. Suppose the state of $\varT$ changes to $\rho_{\varT|n}$ if outcome $n$ occurs; this occurs with probability $p_n=\Tr(\rho_{\varT A}\Pi_n)$. The average Von Neumann entropy of system $\varT$ after the measurement is then
\be
S(\varT|A)=
\sum_n p_n S(\rho_{\varT|n}).
\ee
In terms of the reduced state for system $\varT$ we define
\be
J(\varT;A)=S(\rho_{\varT})-\sum_n p_n S(\rho_{\varT|n}).
\ee
This generalizes (\ref{defasym}).
As is fairly standard, we used a different symbol here, $J$, to denote this particular quantum version of the mutual information \cite{ollivier2001}.
$J$ depends on what measurement is performed on $A$.
We may eliminate this dependence  by maximizing $J$ over all  possible measurements.
This optimization, however, is computationally hard \cite{huang2014}.

The symbol $I$ is used for the other quantum definition of mutual information, based on the symmetric classical definition (\ref{defsym}).
This alternative  definition involves no measurements, and reads
\be
I(\varT;A)=S(\rho_\varT)+S(\rho_A)-S(\rho_{\varT A}).
\ee
This is the definition we will use here. 
We note the well-known property that the mutual information between two qubits can equal 2---namely, when their state is maximally entangled---, whereas the classical quantity can be at most 1. This factor of 2 can be interpreted operationally via superdense coding \cite{bennett1992}, which demonstrates how one qubit of an entangled pair can be used to transmit two classical bits of information.

For the QPID, too, we wish to find expressions that do not involve measurements, using the expression (\ref{ZZ}) as classical starting point. We thus need the quantum version of a conditional state.
\subsection{Quantum conditional states}
Classically, we have the conditional probability distribution $\Prob(\varT|A)=\Prob(\varT,A)/\Prob(A)$.
We cannot straightforwardly find a quantum equivalent to this distribution because the two operators $\rho_{\varT A}$ and $\openone_\varT\otimes\rho_A$ do not necessarily commute.
Indeed, there cannot be a conditional state $\rho_{\varT|A}$ such that we always have
\be
S(\rho_{\varT|A})=S(\rho_A)-S(\rho_{\varT A}),
\ee
simply because the quantity  on the right-hand side  may be negative. If it is negative, $\rho_{\varT A}$ is entangled.

Nonetheless, we can define an operator that generalizes $\Prob(\varT|A)=\Prob(\varT,A)/\Prob(A)$ and that has at least some properties in common with a conditional probability distribution. 
Here we choose to define the quantum conditional state as
\be\label{cond}
\rho_{\varT|A}=( \openone_{\varT}\otimes \rho_A^{-1/2})\,
\rho_{\varT A}\, (\openone_{\varT}\otimes \rho_A^{-1/2}).
\ee
This is an operator on the joint system $\varT\otimes A$, but fails to be a density operator in that it may have eigenvalues larger than 1 (and hence its entropy may be negative). This operator can be used to witness bi-partite entanglement \cite{zhang2007}.
The symbol  $\rho_A^{-1/2}$ here is defined straightforwardly on the support of $\rho_A$, and is set to zero on its kernel (thus corresponding to the Moore-Penrose inverse).
One useful perspective on this quantum version of conditional probability distribution can be found in Refs.~\cite{leifer2013,leifer2014}, where it is argued that this is the proper quantum generalization from a Bayesian perspective.  These two references also demonstrate the usefulness of the $*$ product, defined as
\bea
A*B:=B^{1/2}AB^{1/2},
\eea
such that $\rho_{\varT|A}=\rho_{\varT A} * (\rho_A)^{-1}$, where additional indentity operators have been suppressed.

An alternative definition of conditional states was proposed in Refs.~\cite{cerf1997,cerf1999}, 
\be\label{cond2}
\rho'_{\varT|A}=\lim_{n\rightarrow\infty}
\left[ \openone_{\varT}\otimes (\rho_A)^{-1/(2n)} \, \rho_{\varT A}^{1/n} \, ( \openone_{\varT}\otimes \rho_A)^{-1/(2n)}\right]^{n}.
\ee
It can likewise be employed to detect negative conditional entropy \cite{vempati2021} and entanglement \cite{friis2017}. As a test of entanglement it is weaker than that based on (\ref{cond}). 
This is one reason for using  the conditional operator as defined in (\ref{cond}) in our definition of a QPID. The other reason is that the relation (\ref{cond}) between $\rho_{\varT|A}$ and $\rho_{\varT A}$ can be inverted straightforwardly (in fact, by using the $*$ product), but (\ref{cond2}) cannot. We thus use definition (\ref{cond}).

\section{Quantum PID}

We wish to define quantum versions of $Z_{ab}$ as given in (\ref{ZZ}) and  of $\B_1$ as given in (\ref{BB}). We choose here 
\be
Z_{AB}=\frac12\Tr_{\varT}\left( \rho_{\varT|A}^{1/4} \rho_{\varT|B}^{1/2} \rho_{\varT|A}^{1/4}  +\rho_{\varT|B}^{1/4} \rho_{\varT|A}^{1/2} \rho_{\varT|B}^{1/4}\right).
\ee
Here, of course, identities on $A$ and $B$, respectively, have to be inserted in the definitions of $\rho_{\varT|B}$ and $\rho_{\varT|A}$ so as to produce operators on $A\otimes B\otimes \varT$.
In terms of the $*$ product we may rewrite this as
\be
Z_{AB}=\frac12\Tr_{\varT}\left(  \rho_{\varT|B}^{1/2}*\rho_{\varT|A}^{1/2} 
+  \rho_{\varT|A}^{1/2}*\rho_{\varT|B}^{1/2}  \right).
\ee
The operator $Z_{AB}$ is then
used to define
\be\label{Bqqq}
\Bqo=-\Tr_{AB}\left(\rho_{AB} \log_2(Z_{AB})\right)
\ee
by analogy to the classical  quantity $\B_1$. The analogue of $\B_0$ is then
\be
\Bqz=\frac12 |I(\varT;A)-I(\varT;B)|,
\ee
and we replace (\ref{Bmax}) by
\be
\Bq=\max(\Bqz,\Bqo).
\ee
(In the Appendix we numerically compare these two pooling methods
for typical quantum states of three qubits and of three qutrits, with the results that typically (but not always) the more complicated logarithmic pooling method (yielding $\Bqo$) is superior to the trivial method (yielding $\Bqz$).)

We define (non-negative) unique information by analogy to (\ref{unq}),
\bea\label{unqq}
I_{\unq{A}{B}}&=& \Bq +\frac12\left(I(\varT;A)-I(\varT;B) \right),\nonumber\\
I_{\unq{B}{A}}&=& \Bq +\frac12\left(I(\varT;B)-I(\varT;A)\right).
\eea
[Note that we might alternatively use
\be\label{altZ}
Z'_{AB}=\frac12\Tr_{\varT}\left( \rho_{\varT|A}^{1/2} \rho_{\varT|B}^{1/2} +\rho_{\varT|B}^{1/2}\rho_{\varT|A}^{1/2}\right),
\ee
which does not use the $*$ product and thus lacks some of its nice properties.
This alternative expression is used for some numerical examples in the next two subsections to show that certain features found there are insensitive to this choice.]

Finally, note that the definition of $\Bq$, like that of quantum mutual information, is invariant under local unitary transformations of the form $U=U_T\otimes U_A\otimes U_B$. All information measures introduced here are, therefore, invariant under local unitaries as well.
\subsection{Motivating example}
Let us consider two pure \footnote{We could trivially generalize $P_1$ and $P_2$ to mixed states $\rho_1$ and $\rho_2$, and then $\Bq$ simply reduces to $\B$ by construction.} states corresponding to the classical distributions from Tables I and II, defined by taking equal superpositions of the 8 possible 3-party terms appearing there, and let us denote the two states by
$\ket{\Psi_1}$ and $\ket{\Psi_{2}}$, respectively.
Then we first note that the quantum mutual entropies do not distinguish these two states. For example, for both states we have 
$S(X)+S(\varT)-S(\varT,X)=2$ for $X=A$ and for $X=B$, and 
$S(A,B)+S(\varT)-S(\varT,A,B)=4$.
But the quantity $\Bq$ is clearly different for these two states: it equals 2 for $\ket{\Psi_2}$ and 0 for $\ket{\Psi_1}$. (The same values are found for the alternative definition based on (\ref{altZ}).)  
Thus, the quantity $\Bq$ straightforwardly generalizes the classical quantity $\B$ in this particular case.
The quantity $\Bq$  can be twice as large as the corresponding classical quantity $\B$, just as the quantum mutual information can be twice as large as the classical mutual information.  This indicates $\Bq$ contains both classical and quantum unique information.

\subsection{Scrambling}
The three-variable PID can be applied straightforwardly to the Hayden-Preskill model of a black hole as a processor of quantum information \cite{hayden2007}.
In their setup, adapted to our notation, $\varT$ is a reference system that is initially maximally entangled with a specific small subsystem of $A\otimes B$. That subsystem is thrown into a black hole and the information (i.e., the entanglement with $\varT$) is then scrambled among all degrees of freedom of the full system $A\otimes B$.
System $B$ then is emitted by the black hole as Hawking radiation, and Bob, who collects all that radiation, tries to recover the entanglement with $\varT$.
The part that remains inside the black hole corresponds to system $A$.

In our numerical example we choose the reference system $\varT$ to have Hilbert space dimension $D_\varT=4$, and then choose a pure {\em final}  state for $\varT AB$ such that $\varT$ is (still) maximally entangled with $AB$. Specifically, we write
\be
\ket{\Psi}=\frac{1}{\sqrt{D_\varT}}\sum_{n=1}^{D_\varT}  \ket{n}_\varT \ket{\Phi_n}_{AB},
\ee
where $\{\ket{\Phi_n}_{AB}\}$ for $n=1\ldots D_\varT$
are random orthogonal unit vectors on $AB$. This final state mimics the result of scrambling of quantum information, initially located in a small subsystem of $AB$, spreading throughout the entire system $AB$ (as in \cite{hayden2007}).

	\begin{figure}[h!]
		%made with Qmany10.m
	\includegraphics[width=3in]{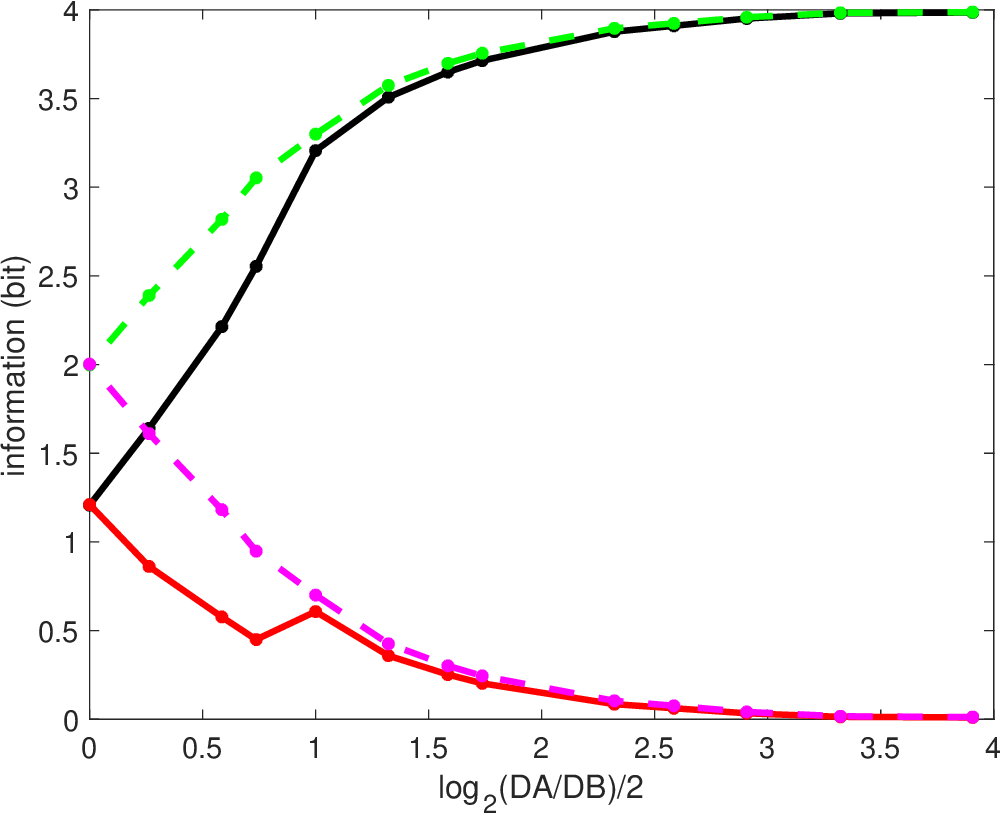}
		\caption{Result of applying a random unitary to system $AB$, starting with a maximally entangled state between $\varT$ and $AB$. Keeping the product of Hilbert space dimensions $D_A$ and $D_B$ fixed and equal to 900 [we could view this system as consisting of 2 qubits, 2 qutrits and 2 5-dimensional systems (ququints)],  we plot unique information for $A$ (solid black upper curve) and for $B$ (solid red lower), as well as  the mutual information $I(\varT;A)$ (dashed green upper curve) and $I(\varT;B)$ (dashed magenta lower curve), all as functions of  $\log_2(D_A/D_B)/2$.
		 The latter can be interpreted as the effective number of qubits that are still part of system $A$ and not yet part of system $B$, counted from the starting point with equal dimensions $D_A=D_B=30$. 
	}
\end{figure}
In the usual setup \cite{hayden2007} $AB$ consists of qubits and one studies how the information quantities $I(\varT;A)$ and $I(\varT;B)$ change when qubits are moved from $A$ to $B$. The ratio $D_A/D_B$ can only change by factors of four in that case. Here we wish to have a more fine-grained description, and we allow qutrits and ququints ($D=5$) as well, thus making possible more values for the ratio $D_A/D_B$. We use $D_{AB}=900=(2\times 3\times 5)^2$ in Figure 1.

We see in that plot that the information about $\varT$ is equally distributed among $A$ and $B$ when they have the same dimension (as it should). 
The green curve gives $I(\varT;A)$, which quickly approaches the maximum information (4 bits) available in systems $A$ and $B$ together as the dimension of system $A$ becomes larger than half the total size. 
Two points illustrated in Figure 1 are worth mentioning.  First,  when the sizes of $A$ and $B$ are the same, more than half of the information each system possesses is unique. Second, once just one qubit has moved from one system to the other (so that the ratio of their dimensions is 4) almost all information possessed by either system is unique.

Note that the unique information in the smaller subsystem tends to decrease with decreasing size, except around the point where the ratio of two Hilbert space dimensions is almost 4 (i.e., starting from equal dimensions we have moved one qubit from one to the other system). At that point,   swapping a qubit and and qutrit  from $A$ for a ququint from $B$ slightly decreases the unique information in $B$ even though its dimension increases.
For the larger subsystem the amount of unique information always increases with increasing dimension, but there is an inflection point clearly visible in the black curve when the ratio of the dimensions is 4. We checked this behavior for a dimension of $D_{AB}=1764$ (see Figure 2), with the same result.
All these conclusions hold as well for the alternative measure based on (\ref{altZ}), as shown in Figure 3.
	\begin{figure}[h!]
	\includegraphics[width=3in]{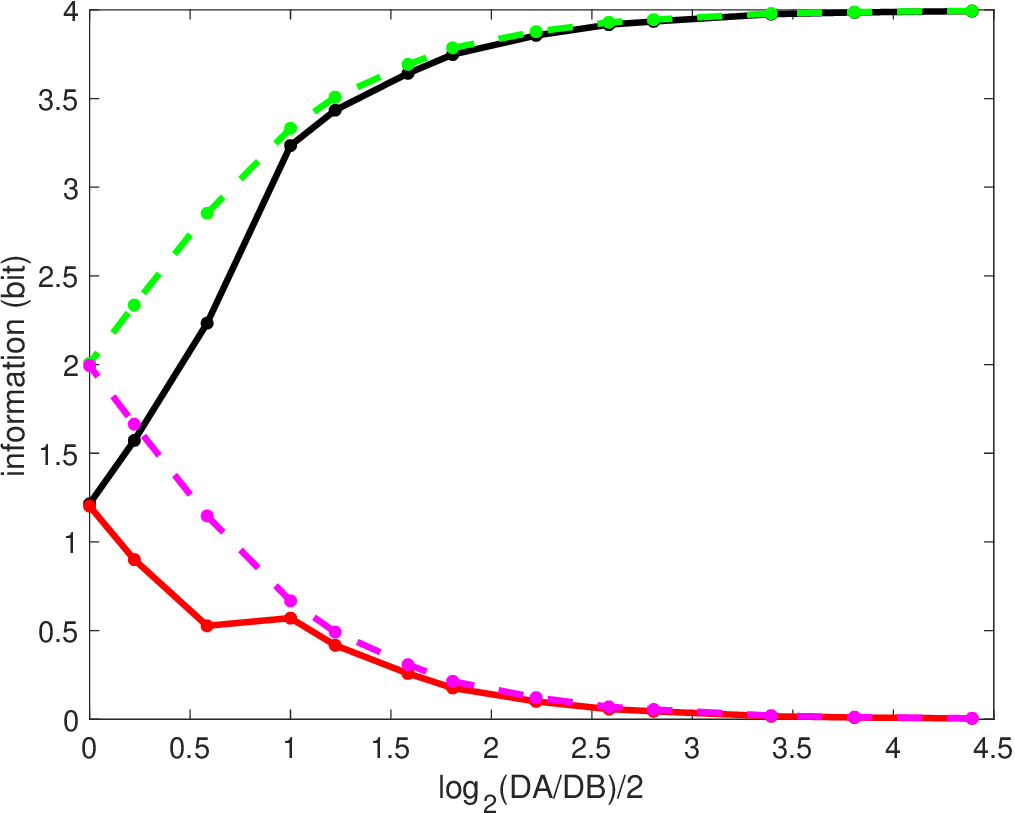}
	\caption{Same as Figure 1, but using $D_{AB}=(2\times 3\times 7)^2=1764$.}
\end{figure}

	\begin{figure}[h!]
	\includegraphics[width=3in]{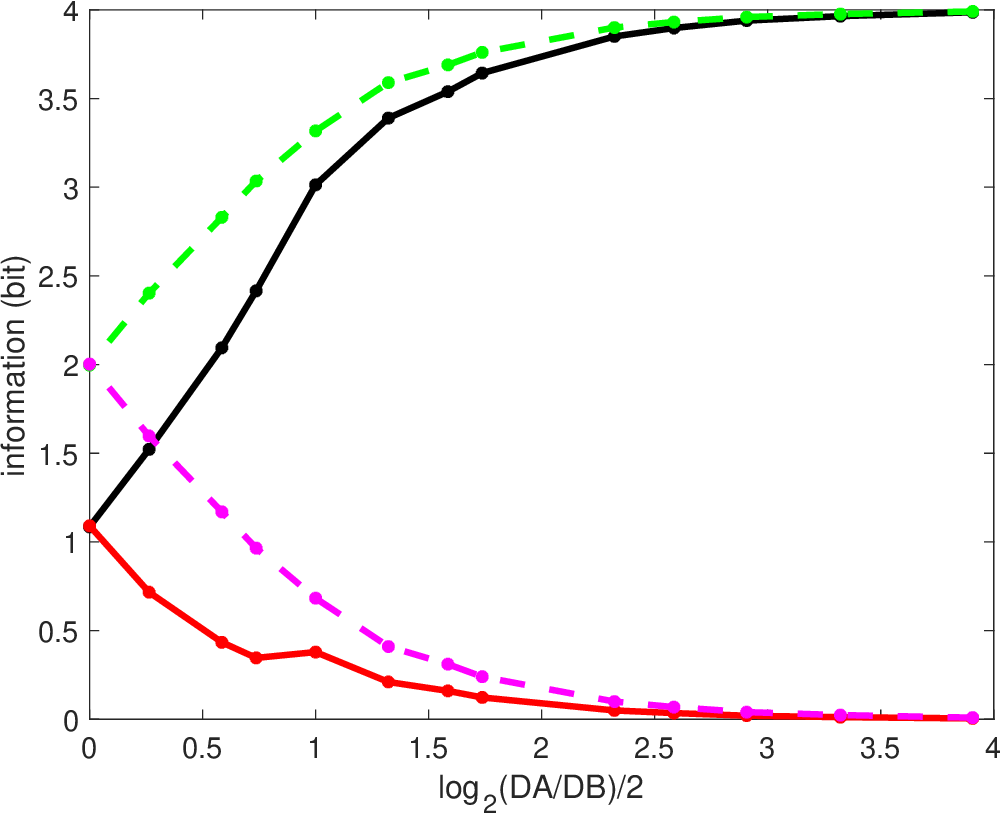}
	\caption{Same as Figure 1, but using the alternative expression (\ref{altZ}) for $Z'_{AB}$.}
\end{figure}

\subsection{Emerging classicality}
One outstanding problem of quantum mechanics is to explain how a classical world emerges from a quantum world. A slightly more manageable version of this problem is to explain how classical information arises out of quantum information.
In this context, the main point of Quantum Darwinism \cite{ollivier2001,zwolak2013,unden2019,touil2020,girolami2022} is that classicality (of a system) emerges from quantum mechanics by having many copies of the same information (about that system) spread out over the environment. 
Here we consider this same problem in terms of unique information.  Let us thus re-analyze an example from Ref.~\cite{touil2020}.
System $\varT$ is one qubit, system $A$ consists of $m_A$ qubits and system $B$ consists of $m_B$ qubits. The initial state of $\varT$ is taken to be
$\ket{\psi}_\varT=\sqrt{p}\ket{0}+\sqrt{q}\ket{1}$, with $p+q=1$. Each of the $N:=m_A+m_B$ qubits starts in the state $\ket{0}$ and then interacts with $\varT$ by a unitary two-qubit interaction, described as a C-MAYBE in Ref.~\cite{touil2020}. This interaction transforms the initial state to a final state of the form
\bea
\ket{\Psi}_{\varT,A,B}=\sqrt{p} \ket{0}\otimes \ket{0}^{\otimes m_A} 
\otimes \ket{0}^{\otimes m_B}+\nonumber\\
+  \sqrt{q} \ket{1}\otimes \ket{r}^{\otimes m_A} 
\otimes \ket{r}^{\otimes m_B},
\eea
where $\ket{r}$ is a slightly rotated state
\be
\ket{r}=s\ket{0}+\sqrt{1-s^2}\ket{1},
\ee
where $s$ is a real number fairly close to 1. In this way, each qubit of the environment encodes some partial information about the state of qubit $\varT$, while all environment qubits taken together contain almost the maximum of two bits of  quantum mutual information. This simple intuition is quantified in FIG.1 of Ref.~\cite{touil2020} and also in FIG.~4 below. In both figures there is a clear ``classical'' plateau visible for which the $m_A$ qubits contain about 1 bit of classical information about qubit $\varT$. That is, the quantum mutual information equals about 1 bit for a large range of values of $m_A$. Using our measure for $I_{\unq{A}{B}}$ we also see that that information is not unique to A. Indeed, the remaining $m_B=N-m_A$  qubits contain more or less the same (classical) information. It is only for either a small number of qubits A or for a large number of qubits A (close to $N$) that we see non-classical behavior. For $m_A$ close to $N$ the mutual information is larger than 1 and thus contains quantum information in addition to the 1 bit of classical information, and we also see that that information is unique to A: both the mutual information and the unique information approach the value 2 (and attain it for $m_A=N$).

We recall that the bonus $\B$ quantifies the asymmetry between $A$ and $B$ w.r.t.~their correlations with $\varT$. For $m_A\ll m_B$ that asymmetry is clearly large. This explains why the unique information in A is nonzero even when $m_A$ is small. It may even be (slightly) larger than the mutual information $I(\varT;A)$, which, as mentioned at the end of Section II can occur for classical probability distributions as well. In FIG.~4 this occurs only when A consists of just one qubit.
	\begin{figure}[h!]
	\includegraphics[width=3in]{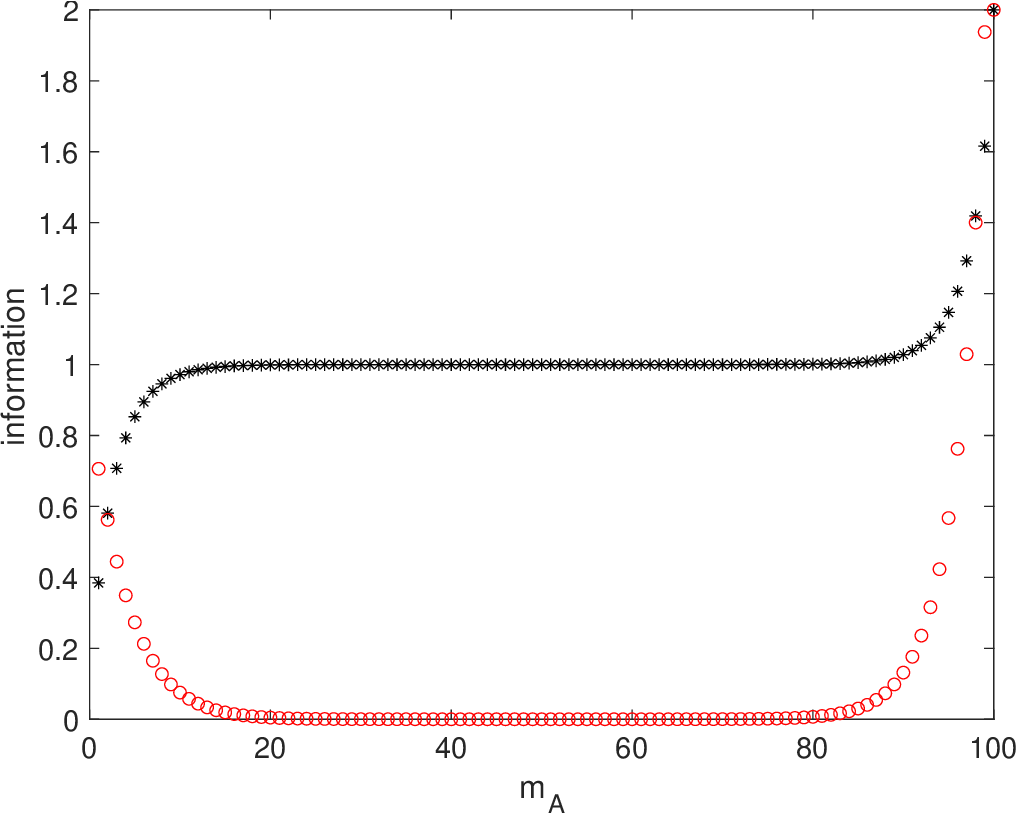}
	\caption{Mututal information $I(\varT;A)$ (stars) and unique information $I_{\unq{A}{B}}$ (circles) plotted as functions of $m_A=1\ldots N$ for a total of $N=100$ qubits. The target system is one qubit, system A consist of $m_A$ qubits and system B consists of $N-m_A$ qubits. Here $s=0.85$ and $p=q=1/2$.}
\end{figure}

\section{Discussion and conclusion}
We introduced a quantum version of the Partial Information Decomposition (PID), which defines synergistic, redundant and unique information and which has proven to be a useful extension of Shannon's information theory.
It  allows one to distinguish correlations and entanglement between quantum systems that cannot be distinguished by standard versions of quantum mutual entropy. We applied the concept to quantum scrambling and quantified the idea that, roughly speaking,  the no-cloning theorem forces there to be unique information. We saw this specifically in the Hayden-Preskill model of a black hole. We defined unique information in terms of a non-negative quantity $\Bq$---given by Eq.~(\ref{Bqqq})---which is zero iff there is no unique information. The quantity $\Bq$ also measures the lack of symmetry between two systems w.r.t. their entanglement with a third reference system. That is, symmetry reduces unique information. We note that, likewise, symmetry reduces scrambling \cite{tajima2021,kudler2022}. 

We end with suggesting three possible further applications.
First, the point of Quantum Darwinism is that classicality emerges from quantum mechanics when many copies of the same information exist in the environment  \cite{ollivier2001,zwolak2013,unden2019,girolami2022}. One can quantify this notion in terms of unique information (or rather, the lack of it), as an alternative to the standard analysis in terms of quantum discord (which is the difference between the (quantum) quantities $I$ and $J$, as defined in Section III.A). Some preliminary results in this direction were given in Section VI.C.

Second, a powerful way to analyze and characterize quantum chaos and, more generally, how quantum information propagates in quantum many-body systems
is in terms of OTOCs (Out-of-Time-Order Correlators) \cite{hosur2016,ding2016,iyoda2018,touil2020,xu2022}.
These  are observable quantities, but certain information measures, specifically Renyi entropies, can be measured as well \cite{brydges2019,van2012}.
Such information measures often appear as bounds on OTOCs . Thus, Renyi entropy versions of the quantities defined here may be useful in this context.

Third, given that the classical PID has been applied to understanding classical neural networks, and given that ideas about scrambling can be applied to understanding or describing quantum neural networks \cite{shen2020,marrero2021,sajjan2023}, it seems that the quantum version of the PID could be fruitfully employed to understand quantum neural networks as well. 

\section*{Appendix}
We mention here some numerical results for random states (both pure and mixed) of three qubits and three qutrits, respectively.
The point is to see how often $\Bqo>\Bqz$, that is, how often the nontrivial logarithmic pooling method beats the trivial method of choosing the lowest-entropy distribution.

\begin{figure}[h!]
	\includegraphics[width=3in]{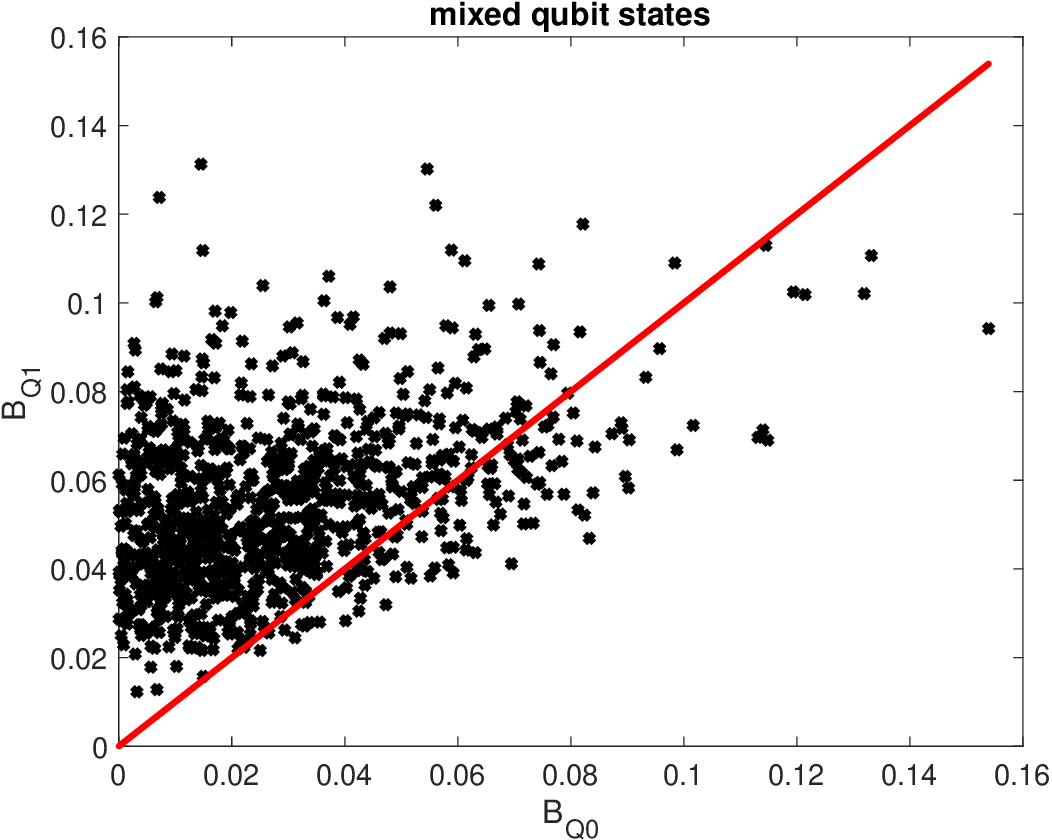}
	\caption{For 1000 randomly chosen mixed states of three qubits, $\Bqo$ vs $\Bqz$, with the red solid line indicating $\Bqo=\Bqz$. For a bit more than 11\% of the states $\Bqo$ fails to be larger than $\Bqz$.}
\end{figure}

	\begin{figure}[h!]
	\includegraphics[width=3in]{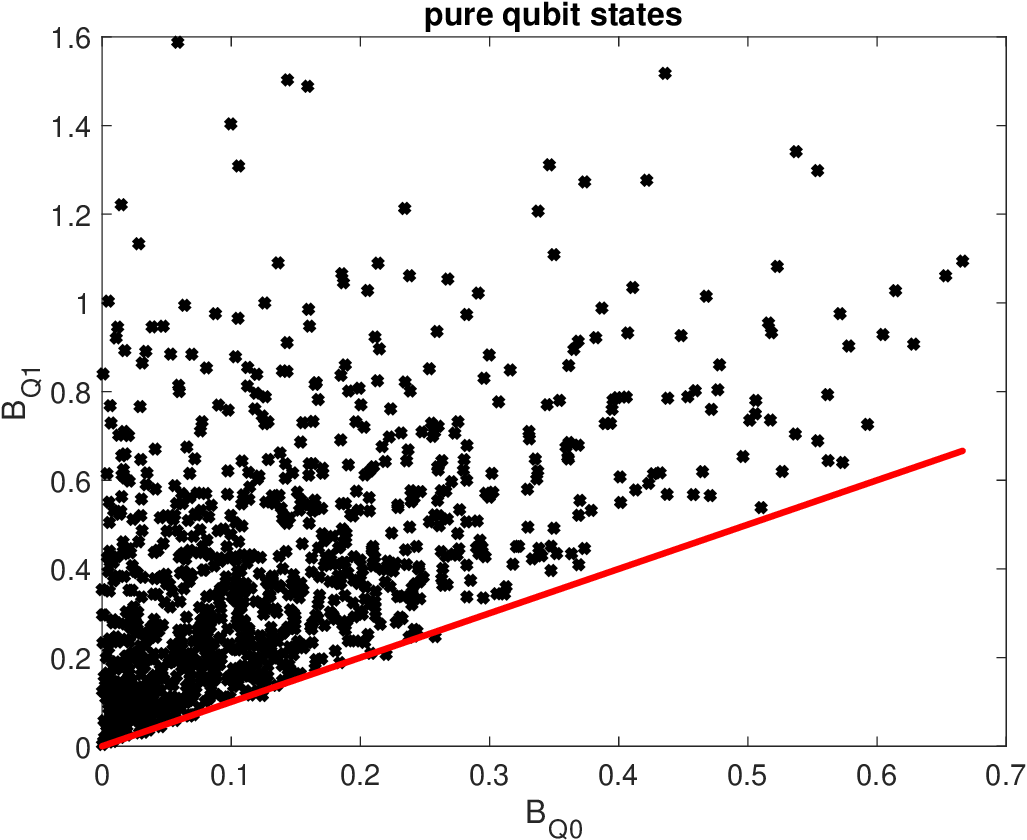}
	\caption{For 1000 randomly chosen pure states of three qubits, $\Bqo$ vs $\Bqz$, with the red solid line indicating $\Bqo=\Bqz$. For slightly less than 1\% of the states $\Bqo$ (just!) fails to be larger than $\Bqz$.}
\end{figure}

For three-qubit states $\rho_{\varT AB}$ we do find 
occasionally that the logarithmic pooling method is worse than simply choosing the lower entropy distribution. This is true for a small fraction (11.4\%, from a sample of $10^5$) of random mixed states, and also for a smaller fraction of random pure states (0.83\% from a sample of $10^5$). See Figures 4 and 5 for typical results for samples of size 1000.

For random three-qutrit states $\rho_{\varT AB}$ we found numerically that the logartihmic pooling method almost always is superior to the simple method of choosing the lower-entropy distribution. Even for mixed states none of a sample of $10^5$ states had $\Bqz>\Bqo$.

\bibliographystyle{unsrt} 
\bibliography{QPID2} 

\begin{thebibliography}{10}

\bibitem{williams2010}
Paul~L Williams and Randall~D Beer.
\newblock Nonnegative decomposition of multivariate information.
\newblock {\em arXiv preprint arXiv:1004.2515}, 2010.

\bibitem{Note1}
For two variables Shannon's mutual information \cite {shannon1948,cover1999}
  suffices.

\bibitem{james2017}
Ryan~G James and James~P Crutchfield.
\newblock Multivariate dependence beyond shannon information.
\newblock {\em Entropy}, 19(10):531, 2017.

\bibitem{Note2}
Co-information is defined as $I({\protect \cal T};A,B)-I({\protect \cal
  T};A)-I({\protect \cal T};B)$. It is also known as interaction information
  \cite {mcgill1954}, and has also been denoted as synergy \cite
  {brenner2000,schneidman2003,latham2005}. In terms of the PID, co-information
  equals the difference between synergistic and redundant information \cite
  {williams2010}.

\bibitem{battiston2021}
Federico Battiston, Enrico Amico, Alain Barrat, Ginestra Bianconi, Guilherme
  Ferraz~de Arruda, Benedetta Franceschiello, Iacopo Iacopini, Sonia K{\'e}fi,
  Vito Latora, Yamir Moreno, et~al.
\newblock The physics of higher-order interactions in complex systems.
\newblock {\em Nature Physics}, 17(10):1093--1098, 2021.

\bibitem{rosas2019}
Fernando~E Rosas, Pedro~AM Mediano, Michael Gastpar, and Henrik~J Jensen.
\newblock Quantifying high-order interdependencies via multivariate extensions
  of the mutual information.
\newblock {\em Physical Review E}, 100(3):032305, 2019.

\bibitem{tax2017}
Tycho~MS Tax, Pedro~AM Mediano, and Murray Shanahan.
\newblock The partial information decomposition of generative neural network
  models.
\newblock {\em Entropy}, 19(9):474, 2017.

\bibitem{wibral2017a}
Michael Wibral, Conor Finn, Patricia Wollstadt, Joseph~T Lizier, and Viola
  Priesemann.
\newblock Quantifying information modification in developing neural networks
  via partial information decomposition.
\newblock {\em Entropy}, 19(9):494, 2017.

\bibitem{wibral2017b}
Michael Wibral, Viola Priesemann, Jim~W Kay, Joseph~T Lizier, and William~A
  Phillips.
\newblock Partial information decomposition as a unified approach to the
  specification of neural goal functions.
\newblock {\em Brain and cognition}, 112:25--38, 2017.

\bibitem{newman2022}
Ehren~L Newman, Thomas~F Varley, Vibin~K Parakkattu, Samantha~P Sherrill, and
  John~M Beggs.
\newblock Revealing the dynamics of neural information processing with
  multivariate information decomposition.
\newblock {\em Entropy}, 24(7):930, 2022.

\bibitem{lizier2018}
Joseph~T Lizier, Nils Bertschinger, J{\"u}rgen Jost, and Michael Wibral.
\newblock Information decomposition of target effects from multi-source
  interactions: Perspectives on previous, current and future work.
\newblock {\em Entropy}, 20(4):307, 2018.

\bibitem{bertschinger2014}
Nils Bertschinger, Johannes Rauh, Eckehard Olbrich, J{\"u}rgen Jost, and Nihat
  Ay.
\newblock Quantifying unique information.
\newblock {\em Entropy}, 16(4):2161--2183, 2014.

\bibitem{james2018b}
Ryan~G James, Jeffrey Emenheiser, and James~P Crutchfield.
\newblock Unique information and secret key agreement.
\newblock {\em Entropy}, 21(1):12, 2018.

\bibitem{harder2013}
Malte Harder, Christoph Salge, and Daniel Polani.
\newblock Bivariate measure of redundant information.
\newblock {\em Physical Review E}, 87(1):012130, 2013.

\bibitem{griffith2014}
Virgil Griffith and Christof Koch.
\newblock Quantifying synergistic mutual information.
\newblock In {\em Guided self-organization: inception}, pages 159--190.
  Springer, 2014.

\bibitem{ince2017}
Robin~AA Ince.
\newblock Measuring multivariate redundant information with pointwise common
  change in surprisal.
\newblock {\em Entropy}, 19(7):318, 2017.

\bibitem{finn2018}
Conor Finn and Joseph~T Lizier.
\newblock Pointwise partial information decomposition using the specificity and
  ambiguity lattices.
\newblock {\em Entropy}, 20(4):297, 2018.

\bibitem{kolchinsky2022}
Artemy Kolchinsky.
\newblock A novel approach to the partial information decomposition.
\newblock {\em Entropy}, 24(3):403, 2022.

\bibitem{rosas2020}
Fernando~E Rosas, Pedro~AM Mediano, Borzoo Rassouli, and Adam~B Barrett.
\newblock An operational information decomposition via synergistic disclosure.
\newblock {\em Journal of Physics A: Mathematical and Theoretical},
  53(48):485001, 2020.

\bibitem{vanenk2023}
Steven~J van Enk.
\newblock Pooling probability distributions and partial information
  decomposition.
\newblock {\em Physical Review E}, 107(5):054133, 2023.

\bibitem{brenner2000}
Naama Brenner, Steven~P Strong, Roland Koberle, William Bialek, and Rob R de
  Ruyter~van Steveninck.
\newblock Synergy in a neural code.
\newblock {\em Neural computation}, 12(7):1531--1552, 2000.

\bibitem{schneidman2003}
Elad Schneidman, William Bialek, and Michael~J Berry.
\newblock Synergy, redundancy, and independence in population codes.
\newblock {\em Journal of Neuroscience}, 23(37):11539--11553, 2003.

\bibitem{latham2005}
Peter~E Latham and Sheila Nirenberg.
\newblock Synergy, redundancy, and independence in population codes, revisited.
\newblock {\em Journal of Neuroscience}, 25(21):5195--5206, 2005.

\bibitem{nielsen2010}
Michael~A Nielsen and Isaac~L Chuang.
\newblock {\em Quantum computation and quantum information}.
\newblock Cambridge university press, 2010.

\bibitem{hayashi2016}
Masahito Hayashi.
\newblock {\em Quantum information theory}.
\newblock Springer, 2016.

\bibitem{wilde2013}
Mark~M Wilde.
\newblock {\em Quantum information theory}.
\newblock Cambridge university press, 2013.

\bibitem{watrous2018}
John Watrous.
\newblock {\em The theory of quantum information}.
\newblock Cambridge university press, 2018.

\bibitem{hayden2007}
Patrick Hayden and John Preskill.
\newblock Black holes as mirrors: quantum information in random subsystems.
\newblock {\em Journal of high energy physics}, 2007(09):120, 2007.

\bibitem{dhar2017}
Himadri~Shekhar Dhar, Amit~Kumar Pal, Debraj Rakshit, Aditi Sen, and Ujjwal
  Sen.
\newblock Monogamy of quantum correlations-a review.
\newblock {\em Lectures on General Quantum Correlations and their
  Applications}, pages 23--64, 2017.

\bibitem{shannon1948}
Claude~E Shannon.
\newblock A mathematical theory of communication.
\newblock {\em The Bell system technical journal}, 27(3):379--423, 1948.

\bibitem{cover1999}
Thomas~M Cover.
\newblock {\em Elements of information theory}.
\newblock John Wiley \& Sons, 1999.

\bibitem{renyi1961}
Alfr{\'e}d R{\'e}nyi.
\newblock On measures of entropy and information.
\newblock In {\em Proceedings of the Fourth Berkeley Symposium on Mathematical
  Statistics and Probability, Volume 1: Contributions to the Theory of
  Statistics}, volume~4, pages 547--562. University of California Press, 1961.

\bibitem{Note3}
For example, Ref.~\cite {james2018b} defines a PID in terms of secret key
  rates. One expects ideas from quantum secret sharing \cite {hillery1999} to
  lead to a QPID, too.

\bibitem{bhat1946}
Anil Bhattacharyya.
\newblock On a measure of divergence between two multinomial populations.
\newblock {\em Sankhy{\=a}: the indian journal of statistics}, pages 401--406,
  1946.

\bibitem{lind1988}
Niels~C Lind and Andrzej~S Nowak.
\newblock Pooling expert opinions on probability distributions.
\newblock {\em Journal of engineering mechanics}, 114(2):328--341, 1988.

\bibitem{pool2000}
David Poole and Adrian~E Raftery.
\newblock Inference for deterministic simulation models: the bayesian melding
  approach.
\newblock {\em Journal of the American Statistical Association},
  95(452):1244--1255, 2000.

\bibitem{carvalho2023}
Luiz~M Carvalho, Daniel~AM Villela, Flavio~C Coelho, and Leonardo~S Bastos.
\newblock Bayesian inference for the weights in logarithmic pooling.
\newblock {\em Bayesian Analysis}, 18(1):223--251, 2023.

\bibitem{neyman2022}
Eric Neyman and Tim Roughgarden.
\newblock No-regret learning with unbounded losses: The case of logarithmic
  pooling.
\newblock {\em arXiv preprint arXiv:2202.11219}, 2022.

\bibitem{ollivier2001}
Harold Ollivier and Wojciech~H Zurek.
\newblock Quantum discord: a measure of the quantumness of correlations.
\newblock {\em Physical review letters}, 88(1):017901, 2001.

\bibitem{huang2014}
Yichen Huang.
\newblock Computing quantum discord is np-complete.
\newblock {\em New journal of physics}, 16(3):033027, 2014.

\bibitem{bennett1992}
Charles~H Bennett and Stephen~J Wiesner.
\newblock Communication via one-and two-particle operators on
  einstein-podolsky-rosen states.
\newblock {\em Physical review letters}, 69(20):2881, 1992.

\bibitem{zhang2007}
Zhengmin Zhang and Shunlong Luo.
\newblock Probabilistic interpretation of the reduction criterion for
  entanglement.
\newblock {\em Physical Review A}, 75(3):032312, 2007.

\bibitem{leifer2013}
Matthew~S Leifer and Robert~W Spekkens.
\newblock Towards a formulation of quantum theory as a causally neutral theory
  of bayesian inference.
\newblock {\em Physical Review A}, 88(5):052130, 2013.

\bibitem{leifer2014}
Matthew~S Leifer and Robert~W Spekkens.
\newblock A bayesian approach to compatibility, improvement, and pooling of
  quantum states.
\newblock {\em Journal of Physics A: Mathematical and Theoretical},
  47(27):275301, 2014.

\bibitem{cerf1997}
Nicolas~J Cerf and Chris Adami.
\newblock Negative entropy and information in quantum mechanics.
\newblock {\em Physical Review Letters}, 79(26):5194, 1997.

\bibitem{cerf1999}
Nicolas~J Cerf and Christoph Adami.
\newblock Quantum extension of conditional probability.
\newblock {\em Physical Review A}, 60(2):893, 1999.

\bibitem{vempati2021}
Mahathi Vempati, Nirman Ganguly, Indranil Chakrabarty, and Arun~K Pati.
\newblock Witnessing negative conditional entropy.
\newblock {\em Physical Review A}, 104(1):012417, 2021.

\bibitem{friis2017}
Nicolai Friis, Sridhar Bulusu, and Reinhold~A Bertlmann.
\newblock Geometry of two-qubit states with negative conditional entropy.
\newblock {\em Journal of Physics A: Mathematical and Theoretical},
  50(12):125301, 2017.

\bibitem{Note4}
We could trivially generalize $P_1$ and $P_2$ to mixed states $\rho _1$ and
  $\rho _2$, and then ${\protect \cal B}_{\protect \cal Q}$ simply reduces to
  ${\protect \cal B}$ by construction.

\bibitem{zwolak2013}
Michael Zwolak and Wojciech~H Zurek.
\newblock Complementarity of quantum discord and classically accessible
  information.
\newblock {\em Scientific Reports}, 3(1):1729, 2013.

\bibitem{unden2019}
Thomas~K Unden, Daniel Louzon, Michael Zwolak, Wojciech~Hubert Zurek, and Fedor
  Jelezko.
\newblock Revealing the emergence of classicality using nitrogen-vacancy
  centers.
\newblock {\em Physical review letters}, 123(14):140402, 2019.

\bibitem{touil2020}
Akram Touil and Sebastian Deffner.
\newblock Quantum scrambling and the growth of mutual information.
\newblock {\em Quantum Science and Technology}, 5(3):035005, 2020.

\bibitem{girolami2022}
Davide Girolami, Akram Touil, Bin Yan, Sebastian Deffner, and Wojciech~H Zurek.
\newblock Redundantly amplified information suppresses quantum correlations in
  many-body systems.
\newblock {\em Physical Review Letters}, 129(1):010401, 2022.

\bibitem{tajima2021}
Hiroyasu Tajima and Keiji Saito.
\newblock Universal limitation of quantum information recovery: symmetry versus
  coherence.
\newblock {\em arXiv preprint arXiv:2103.01876}, 2021.

\bibitem{kudler2022}
Jonah Kudler-Flam, Ramanjit Sohal, and Laimei Nie.
\newblock Information scrambling with conservation laws.
\newblock {\em SciPost Physics}, 12(4):117, 2022.

\bibitem{hosur2016}
Pavan Hosur, Xiao-Liang Qi, Daniel~A Roberts, and Beni Yoshida.
\newblock Chaos in quantum channels.
\newblock {\em Journal of High Energy Physics}, 2016(2):1--49, 2016.

\bibitem{ding2016}
Dawei Ding, Patrick Hayden, and Michael Walter.
\newblock Conditional mutual information of bipartite unitaries and scrambling.
\newblock {\em Journal of High Energy Physics}, 2016(12):1--32, 2016.

\bibitem{iyoda2018}
Eiki Iyoda and Takahiro Sagawa.
\newblock Scrambling of quantum information in quantum many-body systems.
\newblock {\em Physical Review A}, 97(4):042330, 2018.

\bibitem{xu2022}
Shenglong Xu and Brian Swingle.
\newblock Scrambling dynamics and out-of-time ordered correlators in quantum
  many-body systems: a tutorial.
\newblock {\em arXiv preprint arXiv:2202.07060}, 2022.

\bibitem{brydges2019}
Tiff Brydges, Andreas Elben, Petar Jurcevic, Beno{\^\i}t Vermersch, Christine
  Maier, Ben~P Lanyon, Peter Zoller, Rainer Blatt, and Christian~F Roos.
\newblock Probing r{\'e}nyi entanglement entropy via randomized measurements.
\newblock {\em Science}, 364(6437):260--263, 2019.

\bibitem{van2012}
Steven~J van Enk and Carlo~WJ Beenakker.
\newblock Measuring tr $\rho^n$ on single copies of $\rho$ using random
  measurements.
\newblock {\em Physical review letters}, 108(11):110503, 2012.

\bibitem{shen2020}
Huitao Shen, Pengfei Zhang, Yi-Zhuang You, and Hui Zhai.
\newblock Information scrambling in quantum neural networks.
\newblock {\em Physical Review Letters}, 124(20):200504, 2020.

\bibitem{marrero2021}
Carlos~Ortiz Marrero, M{\'a}ria Kieferov{\'a}, and Nathan Wiebe.
\newblock Entanglement-induced barren plateaus.
\newblock {\em PRX Quantum}, 2(4):040316, 2021.

\bibitem{sajjan2023}
Manas Sajjan, Vinit Singh, Raja Selvarajan, and Sabre Kais.
\newblock Imaginary components of out-of-time-order correlator and information
  scrambling for navigating the learning landscape of a quantum machine
  learning model.
\newblock {\em Physical Review Research}, 5(1):013146, 2023.

\bibitem{mcgill1954}
William McGill.
\newblock Multivariate information transmission.
\newblock {\em Transactions of the IRE Professional Group on Information
  Theory}, 4(4):93--111, 1954.

\bibitem{hillery1999}
Mark Hillery, Vladim{\'\i}r Bu{\v{z}}ek, and Andr{\'e} Berthiaume.
\newblock Quantum secret sharing.
\newblock {\em Physical Review A}, 59(3):1829, 1999.

\end{thebibliography}
\end{document}